\documentclass[%
 reprint,
 superscriptaddress,preprintnumbers,
 nofootinbib,
 amssymb,
 aps,
]{revtex4-1}

\usepackage[utf8]{inputenc}
\usepackage[colorlinks=true,citecolor=blue,linkcolor=blue]{hyperref}
\usepackage[normalem]{ulem}
\usepackage{amsmath,amssymb,mathtools}
\usepackage{epsfig}
\usepackage{graphicx}               
\usepackage{url}
\usepackage{color,colortbl}
\usepackage{slashed}
\usepackage{cancel}
\usepackage{multirow}
\usepackage{placeins}
\usepackage[dvipsnames]{xcolor}
\usepackage{epstopdf}
\usepackage{soul}
\usepackage[compat=1.1.0]{tikz-feynman}
\usepackage[capitalise, english]{cleveref}
\usepackage{xspace}
\usepackage{fontawesome5}
\usepackage[english]{babel}
\usetikzlibrary{trees}
\usetikzlibrary{decorations.pathmorphing}
\usetikzlibrary{decorations.markings}

\definecolor{red}{rgb}{0.6,.0706,.1373}
\definecolor{blue}{rgb}{0,0.396,0.741}
\newcommand\myshade{80}
\colorlet{mylinkcolor}{violet}
\colorlet{mycitecolor}{violet}
\colorlet{myurlcolor}{violet}

\hypersetup{
  linkcolor  = mylinkcolor!\myshade!black,
  citecolor  = mycitecolor!\myshade!black,
  urlcolor   = myurlcolor!\myshade!black,
  colorlinks = true
}

\allowdisplaybreaks

\newcommand{\qty}[2]{\ensuremath{#1\,\mathrm{#2}}}
\newcommand{\giga}{G}
\newcommand{\mega}{M}
\newcommand{\electronvolt}{eV}
\newcommand{\barn}{b}
\newcommand{\atto}{a}
\newcommand{\femto}{f}
\newcommand{\pico}{p}
\newcommand{\centi}{c}
\newcommand{\meter}{m}
\newcommand{\second}{s}

\newcommand\aNLO{{\sc\small MadGraph5\_aMC@NLO}}

\def\beq#1\eeq{\begin{align}#1\end{align}}

\makeatletter
\providecommand*{\diff}%
  {\@ifnextchar^{\DIfF}{\DIfF^{}}}
\def\DIfF^#1{%
  \mathop{\mathrm{\mathstrut d}}%
    \nolimits^{#1}\gobblespace}
\def\gobblespace{%
  \futurelet\diffarg\opspace}
\def\opspace{%
  \let\DiffSpace\!%
  \ifx\diffarg(%
    \let\DiffSpace\relax
  \else
    \ifx\diffarg[%
      \let\DiffSpace\relax
    \else
        \ifx\diffarg\{%
        \let\DiffSpace\relax
      \fi\fi\fi\DiffSpace}

\keywords{}

\begin{document}

\title{
Collider and astrophysical signatures of light scalars with enhanced 
\texorpdfstring{\boldmath{$\tau$}}{tau} couplings
}

\author{Jorge Alda}
\email{jorge.alda@pd.infn.it}
\affiliation{Dipartimento di Fisica e Astronomia ``Galileo Galilei'', Universit\`a degli Studi di Padova, 
Via F. Marzolo 8, 35131 Padova, Italy}
\affiliation{Istituto Nazionale di Fisica Nucleare (INFN), Sezione di Padova, Via F. Marzolo 8, 35131 Padova, Italy}
\affiliation{Centro de Astropart\'iculas y F\'isica de Altas Energ\'ias (CAPA) Pedro Cerbuna 12, E-50009 Zaragoza, Spain}
\author{Gabriele Levati}
\email{gabriele.levati@pd.infn.it}
\affiliation{Dipartimento di Fisica e Astronomia ``Galileo Galilei'', Universit\`a degli Studi di Padova, 
Via F. Marzolo 8, 35131 Padova, Italy}
\affiliation{Istituto Nazionale di Fisica Nucleare (INFN), Sezione di Padova, Via F. Marzolo 8, 35131 Padova, Italy}
\author{Paride Paradisi}
\email{paride.paradisi@pd.infn.it}
\affiliation{Dipartimento di Fisica e Astronomia ``Galileo Galilei'', Universit\`a degli Studi di Padova, 
Via F. Marzolo 8, 35131 Padova, Italy}
\affiliation{Istituto Nazionale di Fisica Nucleare (INFN), Sezione di Padova, Via F. Marzolo 8, 35131 Padova, Italy}
\author{Stefano Rigolin}
\email{stefano.rigolin@pd.infn.it}
\affiliation{Dipartimento di Fisica e Astronomia ``Galileo Galilei'', Universit\`a degli Studi di Padova, 
Via F. Marzolo 8, 35131 Padova, Italy}
\affiliation{Istituto Nazionale di Fisica Nucleare (INFN), Sezione di Padova, Via F. Marzolo 8, 35131 Padova, Italy}
\author{Nud\v{z}eim Selimovi\'c}
\email{nudzeim.selimovic@pd.infn.it}
\affiliation{Istituto Nazionale di Fisica Nucleare (INFN), Sezione di Padova, Via F. Marzolo 8, 35131 Padova, Italy}


\begin{abstract}
Beyond Standard Model scenarios addressing the flavor puzzle and the 
hierarchy problem generally predict dominant new physics couplings with fermions of the third generation. 
In this Letter, we explore the collider and astrophysical signatures of new light scalar and pseudoscalar particles dominantly coupled to the $\tau$-lepton. 
The best experimental prospects are expected at Belle II through the $e^+e^-\to\tau^+\tau^-\gamma\gamma,\, \tau^+\tau^-\gamma,\, 3\gamma, {\rm mono-}\gamma$ processes, and the $\tau$ anomalous magnetic moment. 
The correlated effects in these searches can unambiguously point toward the underlying new physics dynamics. 
Moreover, we study astrophysics bounds---especially from core-collapse supernovae and neutron star mergers---finding them particularly effective and complementary to collider bounds.
We carry out this program in the well-motivated context of axion-like particles as well as generic CP-even and CP-odd particles, highlighting possible ways to discriminate among them. 
\end{abstract}

\maketitle

\section{Introduction}
\label{sec:intro}

The LHC discovery of a new scalar with mass around \qty{125}{\giga\electronvolt} and properties compatible with those of the 
Higgs boson, provided a convincing confirmation of the Standard Model (SM) description of electroweak symmetry breaking. Whether the scalar sector chosen by Nature is minimal---as in the SM---or extended---as 
in several beyond SM (BSM) scenarios---remains an important open question of particle physics. Models entailing 
light pseudoscalars, generically dubbed axion-like particles (ALPs) \cite{Jaeckel:2010ni,Marsh:2015xka,Irastorza:2018dyq,DiLuzio:2020wdo}, are among the most renowned BSM scenarios with an extended scalar sector.

Interestingly, ALPs may be helpful in answering several open questions in particle physics such as the strong CP 
\cite{Peccei:1977hh,Peccei:1977ur,Weinberg:1977ma,Wilczek:1977pj} and flavor problems \cite{Davidson:1981zd,Wilczek:1982rv,Berezhiani:1989fp,Calibbi:2016hwq,Greljo:2024evt}, the 
evidence of dark matter \cite{Abbott:1982af,Preskill:1982cy,Dine:1982ah,Davis:1986xc}, as well as the stability of the electroweak scale \cite{Graham:2015cka}.
Their lightness, relative to the new physics (NP) scale from which they stem, can be naturally justified if 
they are pseudo-Nambu-Goldstone bosons associated with the spontaneous breaking of an underlying global symmetry.
The QCD axion, originally proposed as a solution to the strong CP problem, stems from the spontaneous breaking 
of a global $U(1)_{\rm PQ}$ symmetry at the scale $f_a$. Non-perturbative QCD effects provide the axion with an 
effective potential at low energy, leading to the condition $m_a f_a \simeq m_\pi f_\pi$. Instead, the more general 
case, where the ALP mass ($m_a$) and the symmetry breaking scale ($f_a$) are independent parameters,
defines the ALP scenario. In this framework, ALP interactions with fermions and gauge 
bosons of the SM are described through effective  dimension-5 operators~\cite{Georgi:1986df}. Such a model-independent 
approach enables us to capture the general features of broader classes of models without relying on specific 
ultraviolet completions.

For masses below the MeV scale, ALPs can be probed by a variety of cosmological and astrophysical experimental searches. 
This vast program spans from searches in the sub-eV region (such as haloscopes~\cite{ADMX:2003rdr,Barbieri:2016vwg,Caldwell:2016dcw}, helioscopes~\cite{Zioutas:1998cc,Irastorza:2011gs, CAST:2017uph,Armengaud:2014gea}, and 
optical/EM setups~\cite{VanBibber:1987rq,Bahre:2013ywa,OSQAR:2015qdv,Arvanitaki:2014dfa}), to beam-dump 
experiments extending up to the GeV scale \cite{Alekhin:2015byh,Dobrich:2015jyk,Shan:2024pcc}. On the other hand, the mass 
window above the MeV scale can be explored at colliders and through a plethora of rare processes~\cite{Bauer:2017ris,Bauer:2021mvw}.

ALP couplings to charged leptons were studied at $B$- and charm-factories \cite{BaBar:2009oxm,BaBar:2012sau,Belle:2021rcl,BaBar:2009lbr,BaBar:2012wey,Belle-II:2020jti,BESIII:2021ges,BESIII:2022rzz,Belle-II:2023ydz,Belle-II:2024wtd}. 
As a result, an ALP decaying into electrons and muons was constrained for masses up to \qty{10}{\giga\electronvolt}. 
Even though it is experimentally difficult to treat the final states with multiple neutrinos, the BaBar and Belle collaborations managed to constrain ALP decays into pairs of $\tau$ leptons~\cite{BaBar:2009oxm,BaBar:2012sau,Belle:2021rcl}. Most recently, the Belle II collaboration reported the search for an ALP decaying into $\tau$ pairs in $e^+e^-\to \mu^+\mu^-\tau^+\tau^-$ events in the $3.6-\qty{10}{\giga\electronvolt}$ 
mass range \cite{Belle-II:2023ydz}.

These searches have been conducted under the assumption that ALP couplings  to leptons, in the derivative basis, 
are universal.
However, several BSM scenarios addressing the flavor puzzle and the hierarchy problem feature dominant couplings to the 
third fermion generations. A famous paradigm is provided by $U(2)$ flavor models which have been employed e.g. within SUSY 
\cite{Barbieri:1995uv,Barbieri:1996ae,Barbieri:2012uh}, composite Higgs \cite{Matsedonskyi:2014iha,Panico:2016ull,Glioti:2024hye}, and non-universal gauge interactions~\cite{Li:1981nk,Froggatt:1978nt,1983PhLB..129...99B,Bordone:2017bld,Fuentes-Martin:2022xnb,Davighi:2023iks} frameworks. 
Therefore, we find it relevant to explore the phenomenological implications of ALPs dominantly coupled to the tau lepton. 
Since it is difficult to access experimentally the $e^+e^-\to \tau^+\tau^- a (\to\tau^+\tau^-)$ channel, 
which is directly sensitive to the ALP-tau coupling,
we will mainly focus on the 
$\gamma a (\to \rm{inv})$, $\gamma a (\to\gamma\gamma)$, 
$\gamma a (\to\tau^+\tau^-)$, and $\tau^+\tau^-a (\to\gamma\gamma)$ 
processes, where the ALP-photon coupling is unavoidably loop-induced 
via the ALP-tau interaction.

Moreover, the $\tau$ anomalous magnetic moment---which is expected to be probed 
at Belle II with an experimental resolution of $\mathcal{O}(10^{-6})$ through measurements of longitudinal and transverse asymmetries in $\tau$-pair events~\cite{Bernabeu:2007rr,Bernabeu:2008ii,Chen:2018cxt,Crivellin:2021spu}---should receive a large contribution once the ALP-$\tau$ interaction is switched on.

The ALP-$\tau$ coupling can receive significant constraints also from astrophysics observables, like for instance  
core-collapse supernovae and neutron star mergers, via its inevitable one-loop contribution to the ALP-photon 
interaction \cite{Brockway:1996yr,Grifols:1996id}.
The aforementioned physics program is carried out in the well-motivated context of ALPs as well as generic CP-even 
and CP-odd particles, with the intention of highlighting specific signatures enabling us to discriminate among them.

The Letter is structured as follows: in Sec.~\ref{sec:ALP_EFT}, 
we introduce the ALP Effective Field Theory (EFT) with enhanced $\tau$ couplings.
In Sec.~\ref{sec:BelleII}, we discuss the constraints on the ALP-$\tau$ lepton coupling from current and future direct searches 
at colliders while, in Sec.~\ref{sec:tau_gm2}, we analyze the future prospects on the $g-2$ of the $\tau$ at Belle II. 
Sec.\ref{subsec:astro} is devoted to the study of astrophysics bounds. 
In Sec.~\ref{sec:cgg0_impact}, we investigate the impact of a 
direct ALP-photon coupling in addition to the loop-induced one. 
In Sec.~\ref{sec:comparison}, we compare the results of the ALP framework with those pertaining to models with generic 
light scalar and pseudoscalar particles. Finally, in Sec.~\ref{sec:conclusion}, we provide our conclusions.

\section{ALP Effective Field Theory}\label{sec:ALP_EFT}
%
The EFT accounting for ALP interactions with the $\tau$ lepton below the electroweak scale, is described by the following 
dimension-5 Lagrangian
\begin{align}
    \label{eq:Lag_eff}
    \mathcal{L}^a_{d\le5} &=\frac{1}{2}(\partial_\mu a)^2- \frac{1}{2} m_a^2\, a^2\nonumber-\frac{\alpha_{\rm em}}{4\pi}c_{\gamma\gamma}^0\frac{a}{f_a}\,F_{\mu\nu}\Tilde{F}^{\mu\nu}\\&
    -\frac{c_{\tau}}{2f_a} (\partial_\mu a)\,\bar{\tau}\gamma^\mu\gamma_5\tau\,,
\end{align}
where $m_a$ is the ALP mass and $f_a$ is the scale at which the global symmetry is broken. 
An anomalous contribution to the ALP-photon coupling is unavoidably generated at one-loop level through the ALP-$\tau$ 
coupling. For the sake of comparison with a broader class of ALP models, we include in Eq.~\eqref{eq:Lag_eff} a ``bare'' 
ALP-photon coupling, $c_{\gamma\gamma}^0$, encoding additional contributions arising from possible heavy states. 
For phenomenological reasons, it is often preferable to switch from the so-called ``derivative'' basis of 
Eq.~(\ref{eq:Lag_eff}) to the ``chirality-flipping'' one in which the ALP interactions read
\begin{align}
    \mathcal{L}_{\rm int}^a &\supset i \,m_\tau\frac{c_\tau}{f_a}\, a\, \bar{\tau}\gamma_5\tau-
    \frac{\alpha_{\rm em}}{4\pi}\,\frac{c_{\gamma\gamma}^0+c_\tau}{f_a} a\,F_{\mu\nu}\Tilde{F}^{\mu\nu}\,.
\end{align}

In this scenario, the ALP can decay either into a pair of photons or into a $\tau$-lepton pair, if kinematically allowed. 

For $m_a<2m_\tau$, the ALP decays into a pair of photons with the rate given by
\begin{equation}
    \Gamma(a \to \gamma \gamma) = \frac{\alpha^2_\mathrm{em}}{(4\pi)^3} m_a^3 \frac{|c_{\gamma\gamma}^{\mathrm{eff},0}|^2}{f_a^2}\,,
    \label{eq:width_photons}
\end{equation}
with
\begin{equation}
    c_{\gamma\gamma}^{{\rm eff},0} = c^0_{\gamma\gamma} + c_\tau\, B_1\!\left(\!\frac{4m_\tau^2}{m_a^2}\!\right)\,.\label{eq:cgamma_eff}
\end{equation}
The loop function $B_1(x)$ is defined in App.~\ref{app:loop_functions} and, in the $x \gg 1$ limit (i.e. $2\, m_\tau 
\gg m_a$), it is well approximated by $B_1(x)\approx - 1/(3x)$. 

Instead, for $m_a > 2m_\tau$, the $a\to \tau^+\tau^-$ 
channel clearly dominates and the corresponding rate reads
\begin{equation}
    \Gamma(a \to \tau^+\tau^-) = \frac{m_a m_\tau^2}{8\pi f_a^2}|c_{\tau}|^2 \sqrt{1-\frac{4 m_\tau^2}{m_a^2}}\,.
    \label{eq:width_taus}
\end{equation}
While our primary focus is on scenarios where new physics couplings are dominantly connected to third-generation fermions, it is unavoidable that couplings to first- and second-generation leptons are induced at the loop level. Indeed, two-loop renormalisation group (RG) effects generate irreducible couplings to muons and electrons even if one assumes that ALP interactions at the new physics scale $f_a$ involve third-generation leptons only. We estimated these effects using leading-log RG contributions derived in~\cite{Bauer:2020jbp} and find that while the branching ratios for ALP decays to $\mu^+\mu^-$ and $e^+e^-$ are nonzero, they remain negligible (always below $10^{-3}$ level) compared to dominant decays into $\tau$-leptons or photons for relevant mass ranges. Therefore, the main conclusions regarding the light-scalar phenomenology derived below remain practically unaffected by the existence of extra decay channels to $\mu^+\mu^-$ and $e^+e^-$.

\section{Collider searches}
\label{sec:BelleII}
Colliders have already set stringent constraints on ALP interactions with photons, muons, and $\tau$-leptons in the 
mass range $m_a\in[0.2,10]$ GeV \cite{BaBar:2009oxm,BaBar:2009lbr,BaBar:2012wey,BaBar:2012sau,Belle-II:2020jti,
Belle:2021rcl,BESIII:2021ges,BESIII:2022rzz,Belle-II:2023ydz,Belle-II:2024wtd}. 

Nevertheless, none of the searches conducted so far has tested or interpreted data in terms of a purely $\tau$-philic 
scenario. Even for the Belle II search of a $\tau^+\tau^-$ resonance~\cite{Belle-II:2024wtd}, the collaboration 
relies on ALPs radiated from muons, making this search inapplicable in our scenario. Therefore, in this letter, we 
propose a way to bridge this gap by looking for correlated signals in the final states with mono-$\gamma$, $3\gamma$, 
$\tau^+\tau^-\gamma$, $\tau^+\tau^-\gamma\gamma$ and the $g-2$ of the $\tau$-lepton. 
\begin{figure*}[t]
    \centering
    \includegraphics[width=0.95\linewidth]{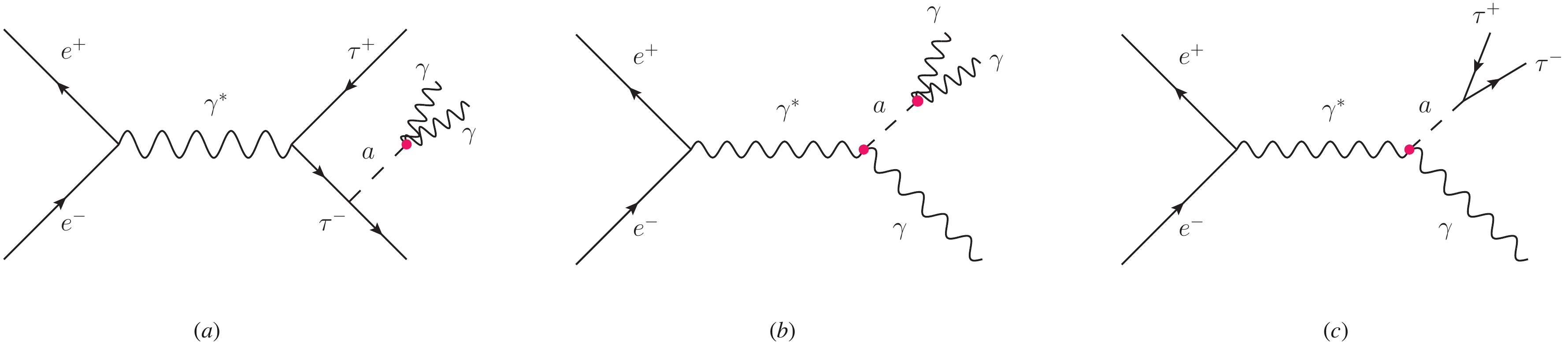}
    \caption{Feynman diagrams for the processes used to directly probe the $\tau$-specific ALP at Belle II. 
    }
    \label{fig:diagrams_direct}
\end{figure*}
In Fig.~\ref{fig:diagrams_direct}, we show the relevant ALP mediated processes at $e^+e^-$ colliders, once 
the Lagrangian in Eq.~\eqref{eq:Lag_eff} is assumed. The $e^+e^- \to 4\tau$ channel is not included here because, 
as previously stated, of difficult experimental implementation. When the ALP is radiated from the final state 
$\tau$-lepton, see Fig.~\ref{fig:diagrams_direct}(a), the production cross-section, at leading order, is proportional 
to $|c_\tau|^2$, and this process is practically insensitive to the value of $c_{\gamma\gamma}^0$, even in the 
$m_a<2m_\tau$ case. 

If, instead, the ALP  production proceeds through an off-shell photon, see Fig.~\ref{fig:diagrams_direct}(b,c), 
the relevant cross-section reads
    \begin{equation}
    \sigma_\mathrm{NR}(e^+e^-\to\gamma a)=\frac{\alpha_\mathrm{em}^3}{24\pi^2}\frac{|c_{\gamma\gamma}^{\mathrm{eff},s}|^2}{f_a^2}\left(1-\frac{m_a^2}{s}\right)^3\,,
    \label{eq:xsec_gammaALP}
\end{equation}
with the effective ALP-photon coupling defined as 
\begin{equation}
    c_{\gamma\gamma}^{{\rm eff},s}=c_{\gamma\gamma}^0 + c_\tau\, B_3\!\left(\!\frac{4m_\tau^2}{m_a^2},
    \frac{4m_\tau^2}{s}\!\right)\,, 
    \label{eq:ALP-offshell_photon}
\end{equation}
hence showing simultaneous sensitivity to $c_{\gamma\gamma}^0$ and $c_\tau$. The loop function $B_3(x,y)$ is defined 
in App.~\ref{app:loop_functions}. It is almost independent of the ALP mass when $m_a \ll 2 m_\tau$, and it introduces an energy dependence of the effective couplings, e.g at BESIII energies $|B_3(x,y)|\approx 0.5$, while $|B_3(x,y)|\approx 1.2$ at Belle II. 

Finally, when the ALP is produced through the decay of a meson resonance $V$, we use the 
Breit-Wigner approximation and write the $e^+e^-\to V \to a \gamma$ cross-section as
\begin{equation}
\!\sigma_{\rm R} = \frac{12\pi\Gamma_V^2}{(s-m_V^2)^2 + m_V^2\Gamma_V^2}\,{\mathcal{B}}(V \!\to\! e^+e^-)
    {\mathcal{B}}(V \!\to\! \gamma a)\,.
    \label{eq:sigmaR_quarkonia}
\end{equation}
Here, $m_V$ and $\Gamma_V$ are the mass and decay width of $V$, $f_V$ is its decay constant~\cite{Merlo:2019anv,DiLuzio:2024jip}, while the branching ratio
$\mathcal{B}(V\to e^+e^-)$ is experimentally determined and can be found in~\cite{ParticleDataGroup:2020ssz}. 
Moreover, the decay of the quarkonium state $V$ to a photon and an ALP is described by the following branching fraction
\begin{equation}
    \mathcal{B}(V \to \gamma a) = \frac{Q_q^2\alpha_\mathrm{em}^3}{24\pi^2 \Gamma_V}m_V f_V^2\frac{|c_{\gamma\gamma}^{\mathrm{eff},s}|^2}{f_a^2}\!\left(\!1-\frac{m_a^2}{m_V^2}\!\right)^{\!3}\!,
    \label{eq:BR_quarkonia}
\end{equation}
with $Q_q$ being the electric charge of the valence quark of the quarkonium. 
By imposing that no evidence for a signature of ALPs decaying to a $\tau^+\tau^-$ or a photon pair is observed, 
we can set limits on $m_\tau |c_\tau| /f_a$ using Eqs.~\eqref{eq:sigmaR_quarkonia} and~\eqref{eq:BR_quarkonia}. 
However, we can only use Eq.~\eqref{eq:BR_quarkonia} directly for resonant searches, i.e. when the parent meson 
has been identified by the kinematics of the process, for example reconstructing the $\Upsilon(1S)$ through 
$\Upsilon(2S)\to \Upsilon(1S) \pi^+ \pi^-$. Instead, if the experiment runs at the energy $\sqrt{s} = m_V$, but the 
meson is not identified kinematically, then the search is sensitive to both non-resonant (Eq.~\eqref{eq:xsec_gammaALP}) 
and resonant (Eq.~\eqref{eq:sigmaR_quarkonia}) cross-sections~\cite{Merlo:2019anv}. In the case of the Belle II 
experiment running at the mass of $\Upsilon(4S)$, the resonance has a width much larger than the energy spread 
of the beam and, consequently, the non-resonant ALP production described by the cross-section in 
Eq.~\eqref{eq:xsec_gammaALP} largely dominates over the resonant contribution.\newline

$e^+e^-\to \tau^+\tau^-\gamma\gamma$: We start with the process in Fig.~\ref{fig:diagrams_direct}(a) and perform a sensitivity study using 
{\sc\small{FeynRules}}-{\sc\small{UFO}}-{\aNLO} chain~\cite{Alloul:2013bka,Degrande:2011ua,Alwall:2011uj} 
to simulate the signal and background events. The signal events consist of the production of an on-shell ALP, 
$e^+e^-\to \tau^+\tau^-a$, that decays into a pair of photons. Therefore, our analysis relies on searching 
for a narrow peak in the photon-pair invariant mass distribution, $m_{\gamma\gamma}^2=m_a^2$, superimposed 
over the smooth QED background. To extract the limits on $|c_\tau|/f_a$, we employ the $m_{\gamma\gamma}^2$ 
resolution reported by Belle II in the $e^+e^-\to \gamma a(\to\gamma\gamma)$ search, where the ALP is 
analogously reconstructed through the two 
recoiling photons~\cite{Belle-II:2020jti}. Furthermore, we require photons with energies $E_\gamma> \qty{1}{\giga\electronvolt}$ in the 
calorimeter region characterized by the polar angle $37.3^{\circ}  < \theta_\gamma < 123.7^{\circ}$, in order to 
have the best energy resolution and minimize the beam background levels~\cite{Belle-II:2018jsg}. 
Likewise, the background from photon conversions outside of the tracking detectors is reduced by requiring angular 
separation between photons $\Delta\theta_{\gamma\gamma} > 0.014$~rad and $\Delta\phi_{\gamma\gamma} > 0.400$~rad 
\cite{Belle-II:2020jti}. Ultimately, we require that at least one of the $\tau$ decays leptonically. We explore 
the ALP mass range $m_a\in[0.4,\,3.5]$ GeV by analyzing the individual $m_{\gamma\gamma}^2$ bins. Utilizing 
Poisson statistics, we determine the upper limit on $m_\tau|c_\tau|/f_a$ for which $S/\sqrt{B}=2$, where $S$ 
represents the number of signal events and $B$ denotes the number of background events in each bin. 
\begin{figure*}[t]
        \centering
        \includegraphics[width=\linewidth]{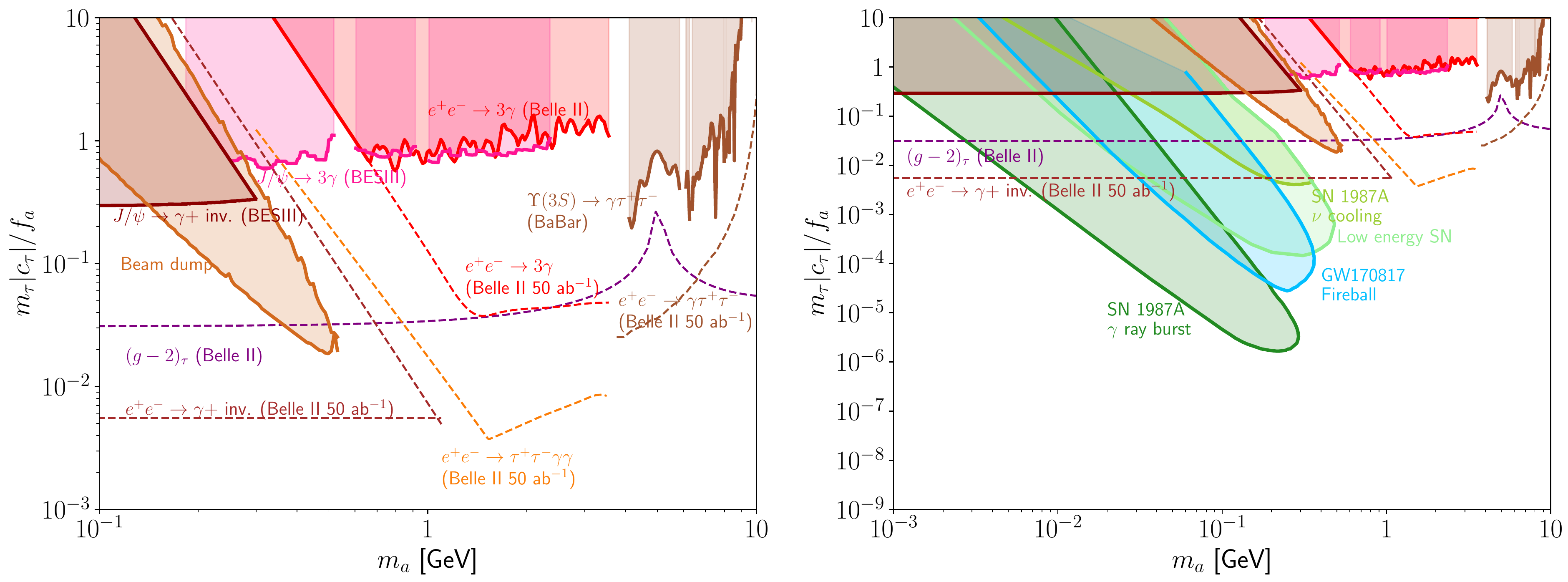}
        \caption{Current (solid lines) and projected (dashed lines) exclusion regions for a $\tau$-specific ALP. Left: Bounds imposed by collider searches, analyzed in Sections~\ref{sec:BelleII} and \ref{sec:tau_gm2}. Right: Bounds arising from both colliders and astrophysics, discussed in Section~\ref{subsec:astro}. All the bounds assume $c_{\gamma\gamma}^0=0$.}
        \label{fig:ALP_summary}
\end{figure*}

At present, no experimental analysis has been performed for this channel. Our projected limits at 95\% CL, 
and for the Belle II foreseen luminosity of \qty{50}{\atto\barn^{-1}}, are shown as an orange-dashed line in the left plot 
of Fig.~\ref{fig:ALP_summary}. This channel has an obvious cutoff at $m_a > 2 m_\tau$ when the $4 \tau$ process 
becomes dominant.
The sensitivity loss for $m_a < \qty{1}{\giga\electronvolt}$ is, instead, due 
to the increased lifetime of lighter ALPs and has been estimated using the Belle search for a leptophilic scalar in the 
$e^+ e^- \to \tau^+\tau^- a(\to \ell^+\ell^-)$ (with $\ell=e,\mu$) channels~\cite{Belle:2022gbl}.
Since our proposal overlaps with this Belle search 
we expect the same sensitivity loss when the ALP proper decay length approaches $L_0\sim \qty{10}{\centi\meter}$. Further investigation is required to determine whether the ALP mass window could be extended through a displaced vertex search.\newline

$e^+e^-\to \tau^+\tau^-\gamma$: In order to explore ALPs with $m_a>2m_\tau$, we use the processes where an ALP decays to a pair of $\tau$-leptons. The searches with $\Upsilon(1S)$ at BaBar~\cite{BaBar:2012sau} and Belle~\cite{Belle:2021rcl} have been performed with data sets of \qty{14}{\femto\barn^{-1}} and \qty{25}{\femto\barn^{-1}}, respectively, while the $\Upsilon(3S)$ at BaBar~\cite{BaBar:2009oxm} used \qty{25}{\femto\barn^{-1}}. As Belle II will mostly run at the energy corresponding to $\Upsilon(4S)$ mass, the mentioned searches with lighter quarkonia will not be further improved. Nevertheless, there is still a bright prospect for testing $\tau$-specific ALPs above the tau production threshold through the non-resonant process $e^+e^-\to \gamma a(\to \tau^+\tau^-)$ shown in Fig.~\ref{fig:diagrams_direct}(c), which will benefit from the large luminosity expected at Belle II. In order to estimate the potential limits, we employ the same {\sc\small{FeynRules}}-{\sc\small{UFO}}-{\aNLO} chain as before and simulate the SM background dominated by QED. As the $\tau$-invariant mass cannot be reconstructed, we focus on the photon energy $E_\gamma> \qty{0.1}{\giga\electronvolt}$ and look for the peak in the distribution of $m_{\tau\tau}^2 = s- 2\sqrt{s} E_\gamma$ which would correspond to the ALP invariant mass. In order to enhance the tagging, we also require both $\tau$-leptons to decay leptonically as in the analogous BaBar search~\cite{BaBar:2009oxm}. The potential 95\% CL limits on $m_\tau |c_\tau|/ f_a$ for $m_a>2m_\tau$ are shown in Fig.~\ref{fig:ALP_summary}(left) for \qty{50}{\atto\barn^{-1}} by the dashed brown line. The current bounds which involve the searches performed at BaBar using $\Upsilon(3S)\to \gamma a(\to\tau^+\tau^-)$~\cite{BaBar:2009oxm} and BaBar and Belle with $\Upsilon(1S)\to \gamma a(\to\tau^+\tau^-)$~\cite{BaBar:2012sau,Belle:2021rcl} are summarised by the solid brown line in Fig.~\ref{fig:ALP_summary}.\newline

\emph{$e^+e^-\to \gamma\gamma\gamma$ and $e^+e^-\to \gamma+{\rm inv}$}: On the other side of the mass spectrum with $m_a<2m_\tau$, the searches involving three photons or a single photon in the final states have already been performed. Examples are given by the Belle II collaboration using a data set of 
$\qty{445}{\pico\barn^{-1}}$
collected at the $\Upsilon(4S)$ energy~\cite{Belle-II:2020jti}, and by the 
BESIII collaboration using a data set of $\qty{2.568}{\femto\barn^{-1}}$ collected 
at the $J/\Psi$ energy~\cite{BESIII:2021cxx,BESIII:2024hdv}. Utilizing Eq.~\eqref{eq:ALP-offshell_photon}, we translate the upper limits 
on the effective ALP-photon interaction to the effective ALP-$\tau$ one, assuming $c_{\gamma\gamma}^0=0$. 
Present bounds are respectively shown as red and pink regions in Fig.~\ref{fig:ALP_summary}(left). Belle II 
prospects with \qty{50}{\atto\barn^{-1}} luminosity are depicted as a red dashed line.
Different regions in Fig.~\ref{fig:ALP_summary}(left) correspond to different experimental signatures expected in 
the $\tau$-specific ALP scenario. The energy of ALPs produced in $e^+e^-\to\gamma a$ processes is fixed and given by
\begin{equation}
    E_a=\frac{s+m_a^2}{2\sqrt{s}}\,,
    \label{eq:ALP_energy}
\end{equation}
which allows us to compute the ALP boost and analyze distinct detector signatures based on the ALP decay length. In the lab frame, it is given by 
\begin{equation}
    L_\mathrm{lab}^a = \frac{\beta_a\gamma_a}{\Gamma(a\to\gamma\gamma)}
\approx \frac{72 (4\pi)^3 \sqrt{s}m_\tau^4 f_a^2}{|c_\tau|^2\alpha_\mathrm{em}^3 m_a^8}\,,
    \label{eq:lab_length_alp}
\end{equation}
where $\beta_a=v_a/c$ is the speed of the emitted ALP, $\gamma_a$ its Lorentz boost, and in we have taken the limit $m_a \ll 2m_\tau,\,\sqrt{s}$ such that 
$\beta_a\gamma_a \approx E_a/m_a$.
We assume that ALPs with a decay length larger than the detector length, $L_{\rm det} = \qty{3}{\meter}$ for Belle II, decay invisibly, and ALPs with a decay length smaller than \qty{1}{\centi\meter} decay promptly~\cite{Belle:2000cnh,Duerr:2019dmv,Belle-II:2018jsg}. 
This defines the lines of constant decay length, $|c_\tau|m_\tau/f_a \propto m_a^{-4}$, separating the different regions corresponding to distinct collider signatures. As one can see, the BESIII search has a similar sensitivity to the current $3\gamma$ search at Belle II. Moreover, the larger detector length, $L_{\rm det} = \qty{7}{\meter}$, and the smaller ALP boost allow us to probe ALP masses in a range inaccessible to Belle II.

An important message from Fig.~\ref{fig:ALP_summary}(left) is that the Belle II collaboration should target the displaced vertex signals with two photons reconstructing the ALP invariant mass for $m_a \lesssim \qty{1}{\giga\electronvolt}$. Further lowering the ALP mass results in the mono-$\gamma$ signature as the ALP decay length becomes of the detector size. The mono-$\gamma$ search has not been performed yet at Belle II, but we can compare it with the BESIII measurement that we show in Fig.~\ref{fig:ALP_summary}(left)~\cite{BESIII:2020sdo}. A dedicated analysis of the interplay between the ALP decay length and the related signatures at Belle II was performed in~\cite{Dolan:2017osp}. We recast their limits on the effective ALP-photon coupling from the mono-$\gamma$ channel and show them in Fig.~\ref{fig:ALP_summary}(left) assuming $c_{\gamma\gamma}^0=0$. In conclusion, the BESIII collaboration provides the best current bounds in the mono-$\gamma$ channel, which will only be exceeded in future Belle II analyses~\cite{BESIII:2020sdo,BaBar:2017tiz,Belle-II:2018jsg}.

Furthermore, in beam dump experiments, ALPs with $m_a \lesssim \qty{1}{\giga\electronvolt}$ could be produced through the Primakoff effect $\gamma N \to a N$, where $N$ is a heavy nucleus, after which the ALP decays into a pair of photons. In Fig.~\ref{fig:ALP_summary} we show the corresponding constraints from SLAC E137~\cite{Bjorken:1988as} and SLAC E141~\cite{Dobrich:2017gcm}.

Finally, we remark that the limits for tauphilic ALP masses below \qty{10}{\mega\electronvolt} and above $2m_\tau$ are so far obtained through one type of process, $e^+e^-\to \gamma + {\rm inv.}$ and $e^+e^-\to \gamma + a(\to \tau^+\tau^-)$, respectively.

An additional handle on these mass ranges is provided by the $\tau$-lepton magnetic moment which we discuss next.

\section{\texorpdfstring{\boldmath{$\tau$}}{tau} anomalous magnetic moment}
\label{sec:tau_gm2}

The most stringent constraint on the tau $g$$-$$2$ arises from the recent measurement of $\tau$-pair production via photon-photon fusion, $pp\to\gamma\gamma\to\tau^+\tau^-$, by the CMS experiment resulting in $a^{\rm exp}_\tau = 9^{+32}_{-31} \times 10^{-4}$~\cite{CMS:2024qjo}. This result improves the current PDG limit at 95\%~CL of $-0.052 < a^{\rm exp}_\tau < 0.013$~\cite{DELPHI:2003nah} obtained from the total cross-section measurement of $e^+ e^-\to e^+ e^-\tau^+\tau^-$ at LEP2.
Still, the situation is anticipated to greatly improve, as Belle II offers a promising way forward through measurements of longitudinal and transverse asymmetries in $\tau$-pair events~\cite{Bernabeu:2007rr,Bernabeu:2008ii, Chen:2018cxt,Crivellin:2021spu}. The expected experimental resolution of $\mathcal{O}(10^{-6})$, 
together with adequate theoretical control, $a_\tau^{\rm SM} = (117717.1\, \pm\, 3.9)\times 10^{-8}$~\cite{Eidelman:2007sb,Keshavarzi:2019abf}, will probe $\tau$-specific ALP couplings in the mass region which was previously unconstrained.

\begin{figure}[t!]
\centering
	\begin{tikzpicture}[baseline = (a)]
	\begin{feynman}[]
	\vertex (a){$\tau$};
    \vertex[right = 0.585  of a](b);
	\vertex [above right = 0.65 of b] (c) ;
	\vertex [below right= 0.65 of c] (d);
     \vertex(a1)[below = 0.85 of c]{$(a)$};
	\vertex [right= 0.52 of d] (e){$\tau$};
    \vertex [above= 0.82 of c] (f){$\gamma$};
	\diagram*{
		(a) -- [plain] (b),
  		(b) -- [plain] (c),
      	(c) -- [plain] (d),
	  	(d) -- [plain] (e),
        (b) -- [scalar, edge label' = \(a\)] (d),
        (f) -- [photon] (c)
	};
	\end{feynman}
	\end{tikzpicture}
 	\begin{tikzpicture}[baseline = (a)]
	\begin{feynman}[]
	\vertex (a){$\tau$};
    \vertex[right = 0.585  of a](b);
	\node [red, dot, above right = 0.65 of b] (c) ;
	\vertex [below right= 0.72 of c] (d);
    \vertex(a1)[below = 1.15 of c]{$(b)$};
	\vertex [right= 0.52 of d] (e){$\tau$};
    \vertex [above= 0.95 of c] (f){$\gamma$};
	\diagram*{
		(a) -- [plain] (b),
  		(b) -- [scalar, edge label = \(a\)] (c),
      	(c) -- [photon] (d),
	  	(d) -- [plain] (e),
        (b) -- [plain] (d),
        (f) -- [photon] (c)
	};
	\end{feynman}
	\end{tikzpicture}
   	\begin{tikzpicture}[baseline = (a)]
	\begin{feynman}[]
	\vertex (a){$\tau$};
    \vertex[right = 0.585  of a](b);
    \vertex[above = 0.48 of b] (c);
    \vertex[above right = 0.5 of c](d);
    \vertex(a1)[below = 1.15 of d]{$(c)$};
    \vertex[above = 0.45 of d](e){$\gamma$};
    \vertex[below right = 0.5 of d](f);
    \vertex[below = 0.48 of f](g);
    \vertex[right = 0.52 of g](h){$\tau$};
    	\diagram*{
		(a) -- [plain] (b),
        (b) -- [plain] (g),
        (g) -- [plain] (h),
  		(f) -- [scalar, edge label = \(a\)] (g),
        (b) -- [photon, edge label = \(\gamma\)] (c),
      	(d) -- [photon] (e),
	  	(c) -- [plain] (d),
    	  (f) -- [plain] (d),
        (f) -- [plain] (c)
	};
	\end{feynman}
	\end{tikzpicture}
\caption{Feynman diagrams contributing to $a_\tau^{\rm ALP}$.}
\label{fig:pEDMALPDiagrams}
\end{figure}
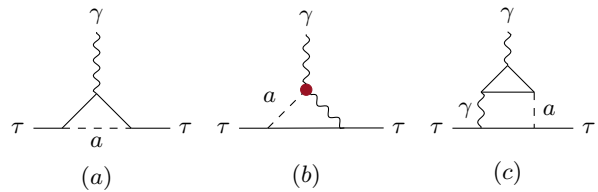

The main contributions to $a^\mathrm{ALP}_\tau \equiv (g-2)_\tau^\mathrm{ALP}/2$ are
\begin{equation}
    a_\tau^\mathrm{ALP} = a_\tau^\mathrm{Yuk} + a_\tau^\mathrm{B-Z} + a_\tau^{\mathrm{ALP-}\gamma}\,,
    \label{eq:g2_ALP}
\end{equation}
where the individual contributions read~\cite{Giudice:2012ms,Marciano:2016yhf,Bauer:2017ris,Cornella:2019uxs}
\begin{align}
\label{eq:gm2}
\!    a_\tau^\mathrm{Yuk} &= 
    -\!\left(\!\frac{m_\tau c_{\tau}}{4\pi f_a}\!\right)^{\!\!2}\!h_1(x_\tau)\,, \\
\!    a_\tau^{\mathrm{ALP-}\gamma} &= m_\tau^2\frac{8\alpha_\mathrm{em}}{(4\pi)^3}\frac{c_\tau}{f_a}
    \frac{c_{\gamma\gamma}^0 \!+ c_\tau}{f_a}\!\!\left(\!
    h_2(x_\tau)-\log\!\frac{\Lambda^2}{m_\tau^2}\!\right)\!,
    \label{eq:ALP-gamma}\\
    a_\tau^\mathrm{B-Z} &=-\!\left(\!\frac{m_\tau c_{\tau}}{4\pi f_a}\!\right)^{\!\!2}\!\frac{2\alpha_\mathrm{em}}{\pi} \!\int_0^1 \!\!dz F(z(1-z)x_{\tau}, x_\tau\!)\label{eq:amu_B-Z},
\end{align} 
and $x_\tau = m_a^2/m_\tau^2$. The scale $\Lambda\sim 4\pi f_a$ signals the ALP EFT breakdown, while we present the functions $h_1$, $h_2$, and $F$ in App.~\ref{app:loop_functions}. The contributions in Eqs.~\eqref{eq:gm2}-\eqref{eq:amu_B-Z} are shown in Fig.~\ref{fig:pEDMALPDiagrams} and are described below:
\begin{itemize}
    \item $a_\tau^\mathrm{Yuk}$ comes from the one-loop diagram
    of Fig.~\ref{fig:pEDMALPDiagrams}(a) and it is always negative.
    \item $a_\tau^{\mathrm{ALP-}\gamma}$ corresponds to the diagram of
    Fig.~\ref{fig:pEDMALPDiagrams}(b). It receives an anomalous contribution which is negative and a contribution from UV physics, encoded in $c_{\gamma\gamma}^0$, which might have either sign.
    \item $a_\tau^\mathrm{B-Z}$ stems from the two-loop Barr-Zee diagram 
    in Fig.~\ref{fig:pEDMALPDiagrams}(c). This contribution is always positive, and for $\tau$-specific ALPs, is always subdominant as compared to the other contributions above.
\end{itemize}

However, the quantity that will be measured at Belle-II with an experimental resolution of $\mathcal{O}(10^{-6})$ is not directly $a^{\rm ALP}_\tau$ 
but rather the $\tau^+\tau^-\gamma^*$ form factor 
$F_2(s)$ at the center-of-mass energy 
$\sqrt{s} \approx 10\,{\rm GeV}$. Our numerical analysis, which is based on the analytical results of~\cite{Cornella:2019uxs}, accounts for the full energy dependence of $F_2(s)$.
In order to have a qualitative understanding of the impact of the energy on the extraction of $a^{\rm ALP}_\tau$ at Belle II, it is useful to consider the limit where $s \gg m_\tau^2, m_a^2$. In this case, $F_2$ is approximated by
\begin{align}
F_2(s \gg m_\tau^2, m_a^2) &= \frac{1}{8\pi^2}\frac{m_\tau^2}{s}\frac{m_\tau^2}{\Lambda^2}c_\tau^2 \,\log \frac{-m_\tau^2}{s} \nonumber \\
& \quad - \frac{\alpha_\text{em}}{4\pi}\frac{c_{\gamma\gamma}\, c_\tau}{4\pi^2}\frac{m_\tau^2}{\Lambda^2} \log \frac{-\Lambda^2}{s}\,.
\end{align}
If the dominant effect to $a^{\rm ALP}_\tau$ arises from 
$a^{\rm Yuk}_\tau$, one finds 
\begin{align}
\label{eq:F2_yuk}
\frac{\mathcal{R}e \, F_2(s)}{a^{\rm ALP}_\tau} 
\approx 
\frac{2m_\tau^2}{s}\,\log\frac{s}{m_\tau^2} \approx 0.2\,,
\end{align}
whereas, if $a^{\rm ALP-\gamma}_\tau$ is dominant, it turns out that
\begin{align}
\label{eq:F2_mix}
\frac{\mathcal{R}e\,F_2(s)}{a^{\rm ALP}_\tau} \approx 
\frac{\log\left(\frac{\Lambda^2}{s}\right)}{\log\left(\frac{\Lambda^2}{m^2_\tau}\right)} \approx \mathcal{O}(1)\,,
\end{align}
where, in the above estimates, we assumed $\sqrt{s} = 10$ GeV.
Notice that, since $a^{\rm ALP-\gamma}_\tau$ is induced by the running of the effective Lagrangian of Eq.~\eqref{eq:Lag_eff}, the corresponding form factor exhibits a logarithmic scaling with the energy. By contrast, $a^{\rm Yuk}_\tau$ 
is finite and the related form factor has a power-law dependence on the energy. 
In Fig.~\ref{fig:ALP_summary}, we show the sensitivity 
of $a^{\rm ALP}_{\tau}$ at Belle II on the ALP parameter space 
(dotted violet line) assuming $c_{\gamma\gamma}^0=0$.

As clearly illustrated by Fig.~\ref{fig:ALP_summary}~(left),
the tau $g$$-$$2$ at Belle II has a unique role in being entirely complementary to searches in all other channels, 
i.e. $\gamma + {\rm inv}$, $3\gamma$, $\tau^+\tau^-\gamma$, and $\tau^+\tau^-\gamma\gamma$. Moreover, a (correlated) signal only in $a_\tau^{\rm ALP}$ and $e^+e^-\to \gamma + {\rm inv}$ or $e^+e^-\to \gamma + a(\to \tau^+\tau^-)$ could be 
identified as the smoking gun of a tauphilic ALP scenario with $m_a \lesssim \qty{10}{\mega\electronvolt}$ or $m_a>2m_\tau$, regions that would be otherwise impossible to access. 

\section{Astrophysical bounds}
\label{subsec:astro}
In astrophysical environments, such as core-collapse supernovae or neutron star mergers, $\tau$-specific ALPs can be generated via the effective coupling to photons of Eqs.~\eqref{eq:cgamma_eff}-\eqref{eq:ALP-offshell_photon} in two different processes: via Primakoff effect, where one real photon is converted into the ALP in the electrostatic field created by the charged particles of the plasma $\gamma + X \to a + X$; and via coalescence of two real photons $\gamma\gamma\to a$. Primakoff effect is the dominant process below $m_a \sim \mathcal{O}(\qty{70}{\mega\electronvolt})$, while coalescence operates up to  $m_a \sim \mathcal{O}(\qty{400}{\mega\electronvolt})$. These ALPs would subsequently decay into a pair of photons, leaving an imprint on several astrophysical events.

\emph{Supernova (SN) $\nu$ cooling}. During the first $\sim\mathcal{O}(\qty{10}{\second})$ after the explosion of the SN 1987A, the proto-neutron star is cooled by the emission of neutrinos. If ALPs or (pseudo)scalars can be produced and efficiently extract energy from the proto-neutron star, the cooling time scale would be significantly shortened. Thus, limits on the effective ALP coupling to photons can be imposed by requiring that the ALP luminosity does not surpass the neutrino luminosity. In the limit of increased ALP-photon couplings, the ALP mean path becomes reduced, and if it becomes smaller than the size of the SN core, in the so-called ``trapping regime'', they can no longer contribute to the cooling process~\cite{Burrows:1990pk,Lucente:2020whw,Caputo:2022rca}.

\emph{No-observation of SN gamma-ray bursts}. If the ALPs produced in a supernova are long-lived, the photons produced in its decay would be observed as a $\gamma$-ray burst. However, in the \qty{223}{s} after the SN 1987A event, the gamma-ray spectrometer (GRS) aboard the Solar Maximum Mission satellite was operational and did not observe said burst~\cite{Hoof:2022xbe,Muller:2023vjm}. The constraints on the ALP coupling are derived by imposing that the ALP has a long enough lifetime so its decay would not have been observed during the operation of the GRS. However, this exclusion is no longer effective if the coupling is so large that the ALP decays inside the envelope of the SN, where the photons would be re-absorbed and would not result in a $\gamma$-ray burst. It should be noted that in part of the parameter space, the photons produced in the decay suffer from additional cooling due to the formation of fireballs (see the text below), and arrive at Earth as X-rays. For this region, GRS is not effective, and exclusion limits are derived from the observations of the Pioneer Venus Orbiter instead~\cite{Diamond:2023scc}.

\emph{Fireballs in neutron star mergers}. The decay of the ALPs produced during a neutron star merger would produce a 
dense plasma of interacting photons dubbed ``fireball''. The fireball undergoes adiabatic and free expansion, such that 
the resulting photons reach Earth with low average energy, where X-ray detectors can detect them. Considering the 
multimessenger signal GW170817/GRB 170817A identified as the asymmetric merger of two neutron stars, the X-ray 
telescopes CALET CGBM, Konus-Wind, and Insight-HXMT/HE can set constraints on the ALP parameters~\cite{Dev:2023hax,Diamond:2023cto}.

\emph{Low-energy supernovae}. If ALPs were short-lived, they would deposit their energy within the progenitor star, contributing to the explosion energy. Studying a population of low-energy supernovae results in bounds complementary to those of gamma-ray bursts~\cite{Caputo:2022mah}. These bounds, similar to those from the diffuse supernova axion-like particle background at lower masses ~\cite{Calore:2021hhn}, are generally more robust than those based on single events like SN1987A and GW170817, which are highly susceptible to systematic uncertainties.

The potential ALP effects in the aforementioned astrophysics processes result in stringent limits on the $\tau$-philic 
ALP couplings, as shown in Fig.~\ref{fig:ALP_summary}(right). These bounds completely dominate the ALP mass region 
$m_a\in[10,400]$ MeV. On the one hand, the limits become weaker for smaller ALP masses due to ALP-photon coupling 
suppression induced by the loop function $B_1$ in Eq.~\eqref{eq:cgamma_eff}, which scales as $B_1\sim -m_a^2/(12m_\tau^2)$ 
for $m_a\ll m_\tau$. On the other hand, for larger ALP masses, astrophysical environments do not have enough energy 
to produce heavier states and become ineffective in constraining $m_a>\mathcal{O}(\qty{100}{\mega\electronvolt})$.
Above these masses, the best prospect comes from colliders' searches at Belle II. Furthermore, we emphasize that the astrophysical constraints on ALP couplings in the trapping regime are subject to significant uncertainties. This is primarily due to the use of perturbative supernova models, which are inadequate in this regime. To highlight this limitation, we represent these constraints with dotted lines, indicating that they should be interpreted with caution.

\section{Impact of \texorpdfstring{\boldmath{$c_{\gamma\gamma}^0$}}{cgg0}
}
\label{sec:cgg0_impact}

\begin{figure}[t!]
    \centering
    \includegraphics[width=0.99\linewidth]{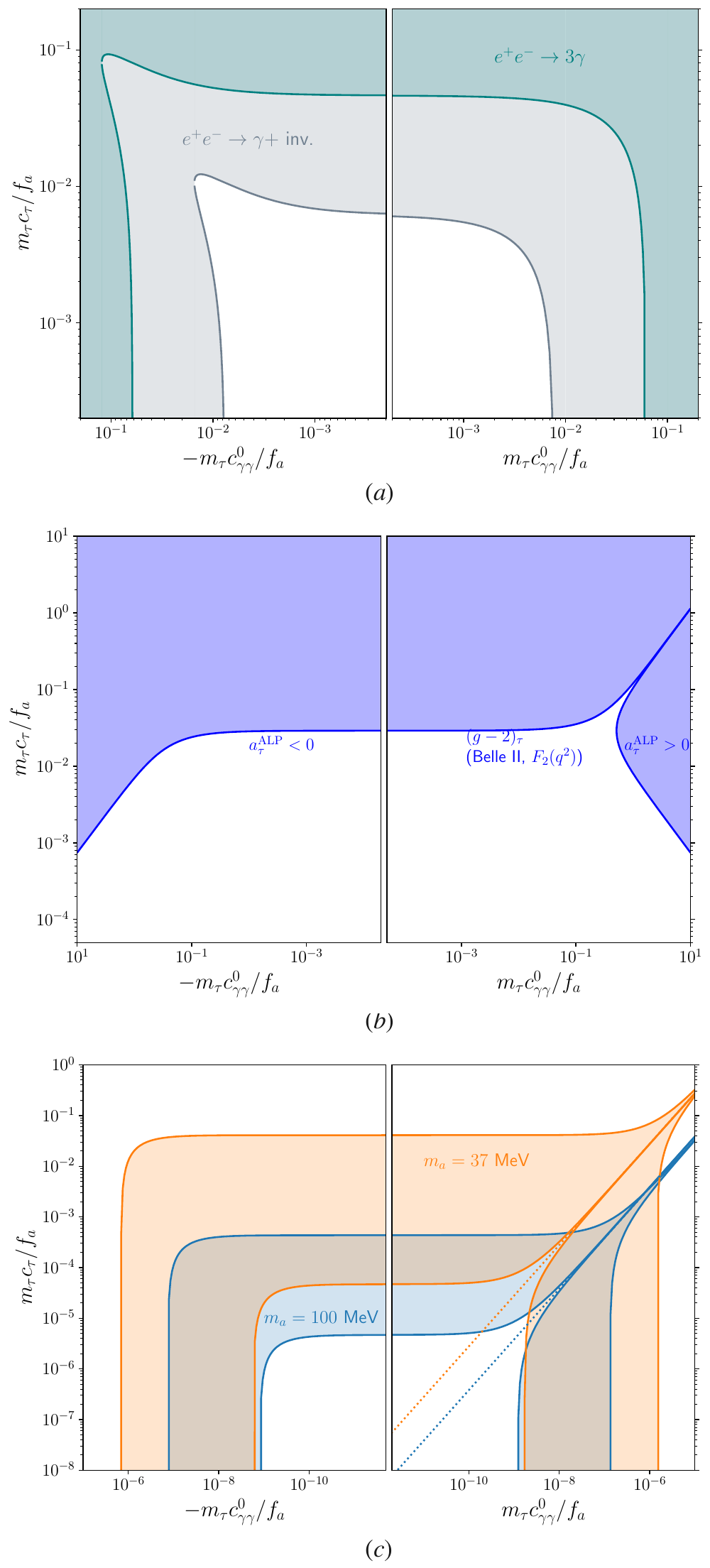}
    \caption{Impact of $c_{\gamma\gamma}^0$ on the $m_\tau c_\tau/f_a$ limits in: \textbf{(a)} The relevant searches at Belle II with maximum luminosity. \textbf{(b)} The anomalous magnetic moment of the $\tau$-lepton. \textbf{(c)} Astrophysics bounds.
    The colored regions are excluded.}
    \label{fig:cgg0_var}
\end{figure}

\begin{figure*}[t!]
    \centering
    \includegraphics[width=\linewidth]{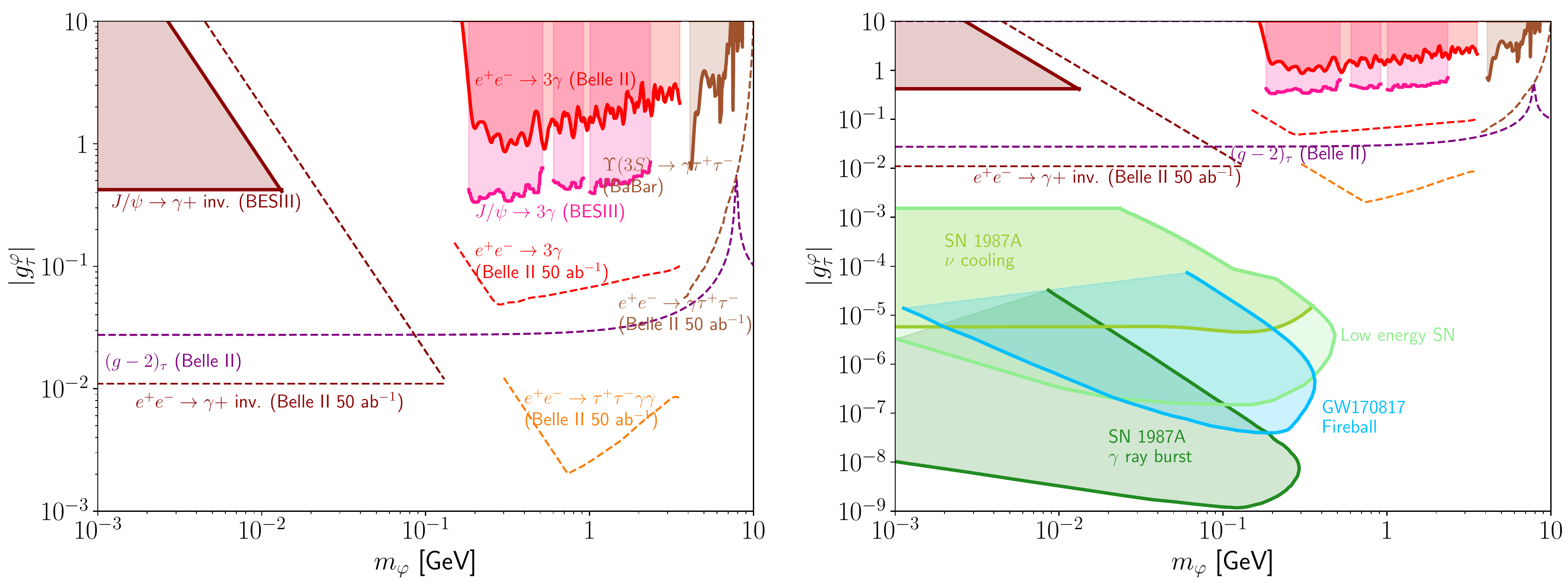}
    \caption{Current (solid lines) and projected (dashed lines) exclusion regions for a $\tau$-specific pseudoscalar. Left: Bounds imposed by collider searches. Right: Bounds arising from both colliders and astrophysics.}
    \label{fig:PS_summary}
\end{figure*}

Up to this point, we assumed that the coefficient $c_{\gamma\gamma}^0$ (Eq.~\eqref{eq:Lag_eff}), which characterizes the effective ALP-photon interactions and is unrelated to the $\tau$-lepton, vanishes. In this section, we describe the consequences of departing from this assumption. In essence, we expect the impact of $c_{\gamma\gamma}^0\ne0$ in all processes which depend on $c_{\gamma\gamma}^{{\rm eff},0}$ and $c_{\gamma\gamma}^{{\rm eff},s}$ defined in Eqs.~\eqref{eq:cgamma_eff} and~\eqref{eq:ALP-offshell_photon}. In our study, these include processes when the ALP is produced together with a photon through $e^+e^-\to \gamma^*\to \gamma a$, the $\tau$ anomalous magnetic moment, as well as astrophysics processes.

We start with the processes at $e^+e^-$ colliders, using Belle II \qty{50}{\atto\barn^{-1}} prospects in mono-$\gamma$ and $3\gamma$ searches to exemplify our findings. The production cross-section $\sigma(e^+e^-\to\gamma a)$ and the ALP decay width are sensitive to $c_{\gamma\gamma}^0$ and its non-vanishing value affects our limits on $m_\tau c_\tau /f_a$. In Fig.~\ref{fig:cgg0_var}(a), we show this interplay by changing the value of $m_\tau c_{\gamma\gamma}^0 /f_a$, and two different signs with respect to $m_\tau c_\tau/f_a$. The gray region is excluded by the $e^+e^-\to\gamma+\rm{inv.}$ search valid for $m_a\lesssim \qty{1}{\giga\electronvolt}$, while the dark green region is excluded by $e^+e^-\to3\gamma$ and is active for $m_a\gtrsim \qty{1}{\giga\electronvolt}$. There are no flat directions when both $c_\tau$ and $c_{\gamma\gamma}^0$ are real, meaning that our bounds are quite robust and only change by a factor of a few for experimentally viable values of $c_{\gamma\gamma}^0$ which do not completely saturate the bounds on $c_{\gamma\gamma}^{{\rm eff},s}$. When $m_\tau |c_{\gamma\gamma}^0|/f_a$ approaches $8\times 10^{-3}$ for mono-$\gamma$ searches and $5\times 10^{-2}$ for $3\gamma$ searches, the bounds on the effective ALP-photon coupling become saturated and limits on $m_\tau c_\tau/f_a$ become stronger.

In the case of the $\tau$-lepton anomalous magnetic moment, a non-vanishing value of $c_{\gamma\gamma}^0$ can induce a change of the $a_\tau^{\rm ALP}$ sign and even a strong suppression of $a_\tau^{\rm ALP}$
due to accidental cancellations between the $a_\tau^{\rm Yuk}$ and $a_\tau^{{\rm ALP}-\gamma}$ terms in Eqs.~\eqref{eq:gm2} and~\eqref{eq:ALP-gamma}. Such cancellation is shown by the flat directions in Fig.~\ref{fig:cgg0_var}(b) when $c_{\gamma\gamma}^0$ and $c_\tau$ have the same sign. We show the excluded regions from Belle II with \qty{50}{\atto\barn^{-1}} in blue. 
In summary, barring unnatural cancellations, the limits on $c_\tau$ from the tau $g$$-$$2$ stay the same or get stronger in the presence of non-zero $c_{\gamma\gamma}^0$.

Lastly, in the case of astrophysics processes, there are again two distinct possibilities based on the relative sign of $c_{\gamma\gamma}^0$ and $c_\tau$. First, if they have the same sign, limits on $m_\tau c_\tau/ f_a$ do not change at all as long as $m_\tau c_{\gamma\gamma}^0/f_a <10^{-8}$, while increasing $c_{\gamma\gamma}^0$ requires $c_\tau$ to lie on a $c_\tau = -c_{\gamma\gamma}^0/B_1(4m_\tau^2/m_a^2)$ line to pass the astrophysics constraints. This situation requires a huge cancellation between two independent parameters and we consider it unnatural. Second, if $c_{\gamma\gamma}^0$ and $c_\tau$ have a different sign, it is impossible to cancel the two contributions and there is no flat direction. Again, if  $m_\tau c_{\gamma\gamma}^0/f_a <10^{-8}$, the bounds on $m_\tau c_\tau/f_a$ remain unchanged, and increasing $c_{\gamma\gamma}^0$ in a range fixed by the data results in stronger limits on $c_\tau$. We exemplify these aspects by bounds based on the non-observation of gamma-ray bursts associated with SN 1987A in Fig.~\ref{fig:cgg0_var}(c). We show how the limits on  $m_\tau c_\tau /f_a$ change as a function of $m_\tau c_{\gamma\gamma}^0/f_a$ for two different ALP masses: the orange region being excluded for $m_a= \qty{37}{\mega\electronvolt}$, and the blue region for $m_a = \qty{100}{\mega\electronvolt}$. The conclusion is that, barring cancellations, the astrophysics bounds we derived either remain the same or get stronger.

\section{Comparison to other scalars}
\label{sec:comparison}

In this section, we compare the phenomenological implications of a $\tau$-specific ALP scenario with the
predictions of models entailing scalars with renormalizable couplings to $\tau$-leptons. We introduce two new states: a pseudoscalar $\varphi$ and a scalar $\phi$ interacting with $\tau$-leptons as follows
\begin{equation}
    \mathcal{L} =i g_\tau^\varphi \varphi\bar{\tau}\gamma_5\tau +  g_\tau^\phi \phi\bar{\tau}\tau\,.
\end{equation}
\begin{figure*}
    \centering
    \includegraphics[width=\linewidth]{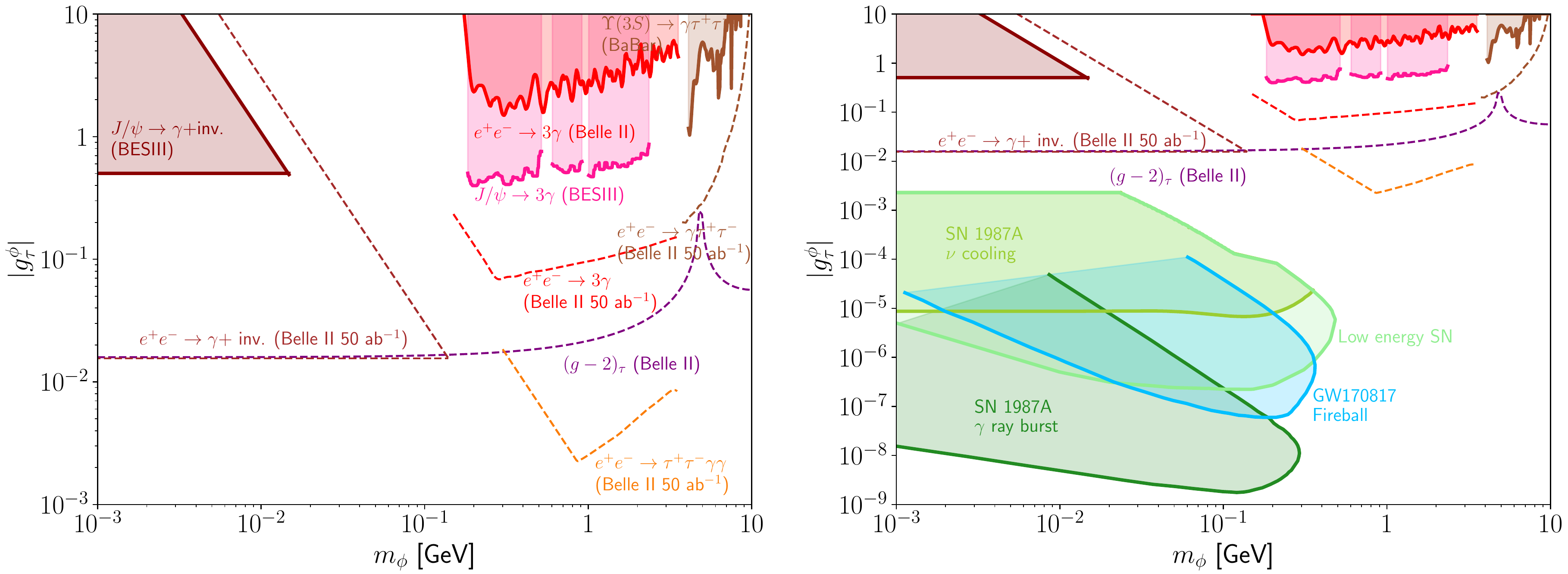}
    \caption{Current (solid lines) and projected (dashed lines) exclusion regions for a $\tau$-specific scalar. Left: Bounds imposed by collider searches. Right: Bounds arising from both colliders and astrophysics.}
    \label{fig:S_summary}
\end{figure*}

At the one-loop level, the following interactions are induced
\begin{equation}
    \mathcal{L} = -g_{\gamma\gamma}^{\varphi} \frac{\alpha_{\text{em}}}{4\pi} \varphi F_{\mu\nu} \tilde{F}^{\mu\nu} - g_{\gamma\gamma}^{\phi}\frac{\alpha_{\text{em}}}{4\pi} \phi F_{\mu\nu} F^{\mu\nu}\,,
\end{equation}
where the effective couplings, with one photon being off-shell, read
\begin{align}
    g_{\gamma\gamma}^{\varphi,s} &= \frac{g_\tau^\varphi}{m_\tau} \left[B_3\!\! \left(\frac{4m_\tau^2}{m_a^2}, \frac{4m_\tau^2}{s}\right)-1\right]\,,\\
    g_{\gamma\gamma}^{\phi,s} & = \frac{g_\tau^\phi}{m_\tau}  A_3\!\! \left(\frac{4m_\tau^2}{m_{\phi}^2}, \frac{4m_\tau^2}{s}\right)\,,
\end{align}
with $s$ being the momentum-squared injected by the off-shell photon, and the loop functions $B_3$ and $A_3$ can be found in App.~\ref{app:loop_functions}.

The effective couplings to photons allow us to derive bounds for $\tau$-specific (pseudo)scalars based on the searches at Belle II and other colliders and the various astrophysical observations, analogously to the ALP case described in the previous section. Interestingly, the distinct loop functions characterizing the spinless particle interactions with photons 
result in significantly different production cross-sections
depending on the CP nature of the particle. For illustration, in the case of Belle II where one of the photons has virtuality $s$ and we take $c_{\gamma\gamma}^0=0$, we find
\begin{align}
       \frac{\sigma(e^+e^-\to\gamma a)}{\sigma(e^+e^-\to\gamma \varphi)} = \frac{|c^{\rm eff,s}_{\gamma\gamma}|^2}{f^2_a |g^{\varphi,s}_{\gamma\gamma}|^2}
       \simeq 4 \times \frac{|c_\tau|^2 m_\tau^2}{|g_\tau^\varphi|^2 f_a^2} 
       \,,
       \label{eq:xsection_ALPvsPS}
\end{align}
for $m_a=m_\varphi<2m_\tau$, and analogously
\begin{align}
\frac{\sigma(e^+e^-\to\gamma \varphi)}{\sigma(e^+e^-\to\gamma \phi)} =
\frac{2}{3} 
\frac{|g^{\varphi,s}_{\gamma\gamma}|^2}{|g^{\phi,s}_{\gamma\gamma}|^2}  
       \simeq 
       12\times \frac{2}{3} \frac{|g_\tau^\varphi|^2}{|g_\tau^\phi|^2}    \,.
\end{align}
This means that the bounds on the ALP-$\tau$ Yukawa coupling $m_\tau |c_\tau|/f_a$ are two times better than the bounds on the pseudoscalar coupling $g_\tau^\varphi$ which is, in turn, roughly two-to-three times more constrained than the scalar coupling  $g_\tau^\phi$. This hierarchy of constraints on the spinless particles is clearly illustrated in Figs.~\ref{fig:ALP_summary},~\ref{fig:PS_summary}, and~\ref{fig:S_summary} in the Belle II searches relying on the ALP production in 
association with a photon.
Furthermore, different loop functions result in spinless particles having distinct decay rates to photons. These are governed by the effective coupling to two on-shell photons $g_{\gamma\gamma}^{S,0}=\lim_{s\to0}g_{\gamma\gamma}^{S,s}$ for $S=\varphi,\phi$. The corresponding ratios read
\begin{align}
    \frac{\Gamma(a\to \gamma\gamma)}{\Gamma(\varphi\to \gamma\gamma)}  = \frac{|c^{\rm eff,0}_{\gamma\gamma}|^2}{f^2_a |g^{\varphi,0}_{\gamma\gamma}|^2}
    \simeq \frac{|c_\tau|^2 m_\tau^2}{|g_\tau^\varphi|^2 f_a^2} \frac{m_a^4}{144 m_\tau^4}\,,\label{eq:photonwidth_aPS}
\end{align}
for $m_a=m_\varphi\ll m_\tau$, and equivalently 

\begin{align}
    \frac{\Gamma(\varphi\to \gamma\gamma)}{\Gamma(\phi\to \gamma\gamma)} = \frac{9}{4}\frac{|g^{\varphi,0}_{\gamma\gamma}|^2}{|g^{\phi,0}_{\gamma\gamma}|^2}
    \simeq \frac{9}{4}\frac{|g_\tau^\varphi|^2}{|g_\tau^\phi|^2}\,.\label{eq:photonwidth_aS}
\end{align}

Consequently, a light $\tau$-specific ALP has a sizably larger lifetime compared to a (pseudo)scalar, meaning that for a fixed mass, it will be experimentally long-lived even for larger couplings. The decay length in the lab frame for a light pseudoscalar is 
\begin{equation}
    \frac{L_\mathrm{lab}^a}{L_\mathrm{lab}^\varphi} = \frac{|g_\tau^\varphi|^2 f_a^2}{|c_\tau|^2 m_\tau^2} \frac{144 m_\tau^4}{m_a^4}\,.
    \label{eq:lab_length_ps}
\end{equation}
An analogous expression for the decay length of the scalar can be obtained from Eq.~\eqref{eq:lab_length_ps} by replacing $|g_\tau^\varphi|^2\to\frac{4}{9}|g_\tau^\phi|^2$ as indicated in Eq.~\eqref{eq:photonwidth_aS}. For both scalar and pseudoscalar, the lines of constant decay length separating visible and invisible decays are of the form $|g_\tau^{\varphi/\phi}|\propto m_a^{-2}$.
This is illustrated when comparing $e+e^-\to \gamma +\rm{inv.}$ or $e+e^-\to 3\gamma$ constraints in Fig.~\ref{fig:ALP_summary} to Figs.~\ref{fig:PS_summary} and~\ref{fig:S_summary}.

More strikingly, the same scaling of the decay rate in Eqs.~\eqref{eq:photonwidth_aPS} and \eqref{eq:photonwidth_aS} explains why the astrophysics bounds are ineffective in the region of very light ALPs, as discussed at the end of Sec.~\ref{subsec:astro}, but are still at work in the case of a generic pseudoscalar or scalar. In addition, we find a remarkable complementarity between the supernovae-based limits, that can not probe large (pseudo)scalar couplings due to the trapping regime kicking in, and future measurements of $(g-2)_\tau$ and the mono-$\gamma$ process at Belle II that would close the astrophysics gap.

Lastly, the modification to the anomalous magnetic moment of the $\tau$ lepton due to a pseudoscalar can be obtained from Eq.~\eqref{eq:g2_ALP} just by setting the anomalous contribution $a_\tau^{\mathrm{ALP-}\gamma}$ equal to zero and replacing $m_\tau c_\tau/f_a \to g_\tau^\varphi$. The effect of a scalar is due to the 1-loop Yukawa-like and 2-loop Barr-Zee contributions
\begin{align}
    a_\tau^{\mathrm{Yuk, }\phi} &= \frac{|g_\tau^\phi|^2}{8\pi^2}\frac{m_\tau^2}{m_\phi^2} I(m_\phi^2/m_\tau^2)\,,\\
    a_\tau^{\mathrm{B-Z, }\phi} &= -\frac{\alpha_\mathrm{em} |g_\tau^\phi|^2}{8\pi^3}\frac{m_\tau^2}{m_\phi^2}L(m_\phi^2/m_\tau^2)\,,
\end{align}
with the loop functions $I$ and $L$ defined in App.~\ref{app:loop_functions}~\cite{Giudice:2012ms}. The Yukawa contribution is the dominant one and it is always positive, while the Barr-Zee term is negative. Phenomenologically, $a_\tau^{\rm ALP}$, $a_\tau^{\varphi}$, and $a_\tau^{\phi}$ give rise to similar constraints on the corresponding particles except for the sign. 

We end this section by noting that the interplay among the different constraints that we presented, together with distinct angular distributions of the photon emitted together with the spinless particle 
\begin{equation}
    \frac{{\rm d}\sigma(e^+e^- \to \gamma a)/{\rm d}\theta}{{\rm d}\sigma(e^+e^- \to \gamma \phi)/{\rm d}\theta} \propto \frac{\sin\theta}{(3+\cos2\theta)}\,,
\end{equation}
could help to probe the CP nature of the associated spin-0 particle at $e^+e^-$ colliders.
\section{Conclusion}
\label{sec:conclusion}
In this Letter, we have explored the collider and astrophysical signatures of new light (pseudo)scalar particles dominantly coupled to the $\tau$-lepton.
This study is motivated by BSM scenarios with dominant couplings to the third fermion family, often invoked as solutions to the flavor and hierarchy problems.

A significant obstacle to probe this scenario through direct searches is the difficulty of reconstructing final states with multiple neutrinos arising from $\tau$-decays.
However, the (pseudo)scalar coupling to the $\tau$-lepton generates the coupling to photons at loop level.
Therefore, we have exploited the direct search processes 
$e^+e^-\to\tau^+\tau^-\gamma\gamma,\,\tau^+\tau^-\gamma,\, 3\gamma,\, {\rm mono-}\gamma$ at colliders. We have proposed new searches at Belle II with $\tau^+\tau^-\gamma\gamma$ and $\tau^+\tau^-\gamma$ final states and derived the corresponding sensitivity limits which we recommend for a more in-depth analysis by the collaboration.

As shown in Fig.~\ref{fig:ALP_summary}, the above mentioned channels 
are very effective and complementary to explore large regions 
of the parameter space of our scenario. 
Moreover, the tau $g$$-$$2$ at Belle II has excellent 
potentialities to probe the (pseudo)scalar parameter space in an entirely complementary way to direct searches. The correlated pattern of new physics effects in these 
observables provides an essential handle on the underlying new physics dynamics. 

Finally, astrophysics bounds from 
core-collapse supernovae and neutron star mergers probe couplings of spin-0 particles to taus up to masses of $\mathcal{O}(\qty{400}{\mega\electronvolt})$, and we find them extremely powerful and complementary to collider bounds.

Our study focused on the well-motivated context of axion-like particles as well as generic CP-even and CP-odd 
particles. Interestingly, as clearly shown in Figs.~\ref{fig:ALP_summary}, \ref{fig:PS_summary}, \ref{fig:S_summary} 
and in Eqs.~\eqref{eq:xsection_ALPvsPS}-\eqref{eq:photonwidth_aS}, the interplay among different constraints that 
we presented, could help us to unveil the CP and/or pseudo-Nambu-Goldstone boson nature of the associated 
spin-0 particle at $e^+e^-$ colliders.

~~~~~~~~~~~~~~~~~~~~~~~~~~~~~~~~~~~~~~~~~~~~~~~~~~~~~~~~~~~~~~~~~~~~~~~~~~~~~

~~~~~~~~~~~~~~~~`

\section*{Acknowledgements}

We thank Edoardo Vitagliano and Sebastian Hoof for useful discussions on astrophysics limits, Arman Korajac for helpful comments about collider searches, Martin Hoferichter for helpful discussions on the tau $g-2$ at Belle II and Enrico Graziani for interesting 
discussions on hidden sectors at Belle II. 
This work received funding by the INFN Iniziativa Specifica APINE and 
from the European Union's Horizon 2020 research and innovation programme 
under the Marie Sk\l{}odowska-Curie grant agreements n. 860881 -- HIDDeN, n.~101086085 -- ASYMMETRY and . This work was also partially supported 
by the Italian MUR Departments of Excellence grant 2023-2027 ``Quantum Frontiers". 
JA has received funding from the Fundaci\'on Ram\'on Areces ``Beca para ampliaci\'on de estudios en el extranjero en el campo de las Ciencias 
de la Vida y de la Materia'', and acknowledges support by the grants PGC2022-126078NB-C21 funded by
MCIN/AEI/10.13039/501100011033.

\appendix
\begin{widetext}
\section{Loop functions}
\label{app:loop_functions}
\subsection{Effective photon couplings}
The loop functions for the effective couplings of spin-0 particles interacting with one on-shell and one off-shell photon, are given by
    \begin{align}
    B_3(x,y) &=\!1\!+\! \frac{xy}{x-y} \left[f^2(x)-f^2(y)\right]\,,\\
     A_3(x,y)&=\frac{xy}{(x-y)^2} \left[x-y +  (x-y+xy)\left(f^2(x)-f^2(y)\right)-x (g(x)-g(y)) \right]\,,
     \end{align}
with
    \begin{align}
    f(x) &= \begin{cases} \arcsin\left(\frac{1}{\sqrt{x}}\right) & x \geq 1 \\ \frac{\pi}{2}+\frac{i}{2}\log \frac{1+\sqrt{1-x}}{1-\sqrt{1-x}} & x<1 \end{cases}\\
    g(x) &= \begin{cases}
        -\sqrt{x-1} \arccos\left(1-\frac{2}{x}\right) & x \geq 1 \\
        \sqrt{1-x} \left(\log\frac{2-x-2\sqrt{1-x}}{x} +i \pi \right) & x<1
    \end{cases}
    \end{align}

In the limit where the two photons are on-shell, $s \ll 4 m_\tau^2$, which is suitable for describing the (pseudo)scalar decay $S\to\gamma\gamma$ (with $S=a\,,\varphi\,,\phi$) as well as its production in supernovae and neutron star mergers, one has
\begin{align}
B_1(x) &= B_3(x,y \gg 1)\simeq\!1\!-\! x f^2(x)\,, \\
B_1(x\gg1) &\simeq -\frac{1}{3x}\,,\\
A_1(x) &= A_3(x, y \gg 1)\simeq-x\left[1\!-\!(x\!-\!1)f^2(x)\right]\,,\\
A_1(x\gg1) &\simeq -\frac{2}{3}\,. 
\end{align}
On the other hand, in the limit $s \gg 4 m_\tau^2$, which is an adequate assumption for processes taking place at Belle-II, one finds the following approximate expressions: 
\begin{align}
B_3(x,y \ll 1)&\simeq\!1\!+\! y \arcsin^2\!\frac{1}{\sqrt{x}} + \frac{y}{4} \log^2\!\left(\!\frac{-y}{4}\!\right), \\
B_3(x\gg1, y\ll1) &\simeq\!1\!+\! \frac{y}{4}\log^2\left(\frac{-y}{4}\right)+\frac{y}{x}\,,\\
A_3(x, y \ll 1) &\simeq y \left[1+f^2(x)-g(x)+\log\frac{-y}{4}+\frac{1}{4}\log^2\frac{-y}{4}\right]\,,\\
A_3(x\gg1, y \ll 1) &\simeq y\left[3+ \!\log \left(\frac{-y}{4}\right) \!+\! \frac{1}{4} \log^2\left(\frac{-y}{4}\right)\right]+ \frac{y}{3x}\,. 
\end{align}
\subsection{Anomalous magnetic moment}
    The loop functions entering the contributions to the anomalous magnetic moment of the $\tau$ are given by
\begin{align}
    h_1(x) &= 1 +2x + x(1-x)\log x -2x(3-x) \sqrt{\frac{x}{4-x}}\arccos\frac{\sqrt{x}}{2}\,,\\
    h_2(x) &= 1 - \frac{x}{3}+\frac{x^2}{6}\log x + \frac{2+x}{3}\sqrt{x(4-x)}\arccos\frac{\sqrt{x}}{2}\,,\\
    F(x, y) &= \frac{1}{1-x}\left[h_2(y)-h_2\left(\frac{y}{x}\right)\right]\,,\\
    I(x) &= \int_0^1 dz\,\, \frac{z^2(2-z)}{1-z+z^2/x}\,,\\
    L(x) &= \int_0^1 dz\,\,\frac{1-2+2z^2}{z-z^2-1/x}\log[(z-z^2)x]\,.
\end{align}
For light ALPs the limits are $h_1(0) \approx h_2(0) \approx 1$, $I(0)\approx \frac{3m_\phi^2}{2m_\tau^2}$, $L(0)\approx \frac{m_\phi^2}{9m\tau^2}(6\log\frac{m_\tau^2}{m_\phi^2}+13)$.
\end{widetext}

\bibliographystyle{JHEP}
\bibliography{references.bib}

\end{document}